\documentclass[aps,prl,reprint,superscriptaddress,showkeys,showpacs]{revtex4-1}
\usepackage[utf8]{inputenc}

\usepackage{graphicx} %
\usepackage{amsmath} %
\usepackage{amssymb} %
\usepackage{subfigure} %
\usepackage{float} %
\usepackage{upgreek} %
\usepackage{multirow} %
\usepackage{color} %
\usepackage{lineno} %
\usepackage{hyperref} %
\usepackage{booktabs}
\usepackage{placeins}
\usepackage[normalem]{ulem}
\usepackage[usenames,dvipsnames]{xcolor}
\hypersetup{colorlinks=true, urlcolor=blue, linkcolor=blue, citecolor=blue} %
\usepackage{lineno}
\usepackage{lipsum}
\usepackage{tabularx}
\usepackage{verbatim}
\usepackage{textgreek}
\usepackage{siunitx}

\begin{document}

\title{Combined Annual Modulation Dark Matter Search with COSINE-100 and ANAIS-112}
\author{N.~Carlin}
\affiliation{Physics Institute, University of S\~{a}o Paulo, 05508-090, S\~{a}o Paulo, Brazil}
\author{J.~Y.~Cho}
\affiliation{Department of Physics, Kyungpook National University, Daegu 41566, Republic of Korea}
\author{J.~J.~Choi}
\affiliation{Center for Underground Physics, Institute for Basic Science (IBS), Daejeon 34126, Republic of Korea}
\author{S.~Choi}
\affiliation{Department of Physics and Astronomy, Seoul National University, Seoul 08826, Republic of Korea} 
\author{A.~C.~Ezeribe}
\affiliation{Department of Physics and Astronomy, University of Sheffield, Sheffield S3 7RH, United Kingdom}
\author{L.~E.~Fran{\c c}a}
\affiliation{Physics Institute, University of S\~{a}o Paulo, 05508-090, S\~{a}o Paulo, Brazil}
\author{C.~Ha}
\affiliation{Department of Physics, Chung-Ang University, Seoul 06973, Republic of Korea}
\author{I.~S.~Hahn}
\affiliation{Center for Exotic Nuclear Studies, Institute for Basic Science (IBS), Daejeon 34126, Republic of Korea}
\affiliation{Department of Science Education, Ewha Womans University, Seoul 03760, Republic of Korea} 
\affiliation{IBS School, University of Science and Technology (UST), Daejeon 34113, Republic of Korea}
\author{S.~J.~Hollick}
\affiliation{Department of Physics and Wright Laboratory, Yale University, New Haven, CT 06520, USA}
\author{S.~B.~Hong}
\affiliation{Department of Physics, Kyungpook National University, Daegu 41566, Republic of Korea}
\author{E.~J.~Jeon}
\affiliation{Center for Underground Physics, Institute for Basic Science (IBS), Daejeon 34126, Republic of Korea}
\affiliation{IBS School, University of Science and Technology (UST), Daejeon 34113, Republic of Korea}
\author{H.~W.~Joo}
\affiliation{Department of Physics and Astronomy, Seoul National University, Seoul 08826, Republic of Korea} 
\author{W.~G.~Kang}
\affiliation{Center for Underground Physics, Institute for Basic Science (IBS), Daejeon 34126, Republic of Korea}
\author{M.~Kauer}
\affiliation{Department of Physics and Wisconsin IceCube Particle Astrophysics Center, University of Wisconsin-Madison, Madison, WI 53706, USA}
\author{B.~H.~Kim}
\affiliation{Center for Underground Physics, Institute for Basic Science (IBS), Daejeon 34126, Republic of Korea}
\author{H.~J.~Kim}
\affiliation{Department of Physics, Kyungpook National University, Daegu 41566, Republic of Korea}
\author{J.~Kim}
\affiliation{Department of Physics, Chung-Ang University, Seoul 06973, Republic of Korea}
\author{K.~W.~Kim}
\affiliation{Center for Underground Physics, Institute for Basic Science (IBS), Daejeon 34126, Republic of Korea}
\author{S.~H.~Kim}
\affiliation{Center for Underground Physics, Institute for Basic Science (IBS), Daejeon 34126, Republic of Korea}
\author{S.~K.~Kim}
\affiliation{Department of Physics and Astronomy, Seoul National University, Seoul 08826, Republic of Korea}
\author{W.~K.~Kim}
\affiliation{IBS School, University of Science and Technology (UST), Daejeon 34113, Republic of Korea}
\affiliation{Center for Underground Physics, Institute for Basic Science (IBS), Daejeon 34126, Republic of Korea}
\author{Y.~D.~Kim}
\affiliation{Center for Underground Physics, Institute for Basic Science (IBS), Daejeon 34126, Republic of Korea}
\affiliation{IBS School, University of Science and Technology (UST), Daejeon 34113, Republic of Korea}
\author{Y.~H.~Kim}
\affiliation{Center for Underground Physics, Institute for Basic Science (IBS), Daejeon 34126, Republic of Korea}
\affiliation{IBS School, University of Science and Technology (UST), Daejeon 34113, Republic of Korea}
\author{Y.~J.~Ko}
\affiliation{Department of Physics, Jeju National University, Jeju 63243, Republic of Korea}
\author{D.~H.~Lee}
\affiliation{Department of Physics, Kyungpook National University, Daegu 41566, Republic of Korea}
\author{E.~K.~Lee}
\affiliation{Center for Underground Physics, Institute for Basic Science (IBS), Daejeon 34126, Republic of Korea}
\author{H.~Lee}
\affiliation{IBS School, University of Science and Technology (UST), Daejeon 34113, Republic of Korea}
\affiliation{Center for Underground Physics, Institute for Basic Science (IBS), Daejeon 34126, Republic of Korea}
\author{H.~S.~Lee}
\affiliation{Center for Underground Physics, Institute for Basic Science (IBS), Daejeon 34126, Republic of Korea}
\affiliation{IBS School, University of Science and Technology (UST), Daejeon 34113, Republic of Korea}
\author{H.~Y.~Lee}
\affiliation{Center for Exotic Nuclear Studies, Institute for Basic Science (IBS), Daejeon 34126, Republic of Korea}
\author{I.~S.~Lee}
\affiliation{Center for Underground Physics, Institute for Basic Science (IBS), Daejeon 34126, Republic of Korea}
\author{J.~Lee}
\affiliation{Center for Underground Physics, Institute for Basic Science (IBS), Daejeon 34126, Republic of Korea}
\author{J.~Y.~Lee}
\affiliation{Department of Physics, Kyungpook National University, Daegu 41566, Republic of Korea}
\author{M.~H.~Lee}
\affiliation{Center for Underground Physics, Institute for Basic Science (IBS), Daejeon 34126, Republic of Korea}
\affiliation{IBS School, University of Science and Technology (UST), Daejeon 34113, Republic of Korea}
\author{S.~H.~Lee}
\affiliation{IBS School, University of Science and Technology (UST), Daejeon 34113, Republic of Korea}
\affiliation{Center for Underground Physics, Institute for Basic Science (IBS), Daejeon 34126, Republic of Korea}
\author{S.~M.~Lee}
\affiliation{Department of Physics and Astronomy, Seoul National University, Seoul 08826, Republic of Korea} 
\author{Y.~J.~Lee}
\affiliation{Department of Physics, Chung-Ang University, Seoul 06973, Republic of Korea}
\author{D.~S.~Leonard}
\affiliation{Center for Underground Physics, Institute for Basic Science (IBS), Daejeon 34126, Republic of Korea}
\author{N.~T.~Luan}
\affiliation{Department of Physics, Kyungpook National University, Daegu 41566, Republic of Korea}
\author{V.~H.~A.~Machado}
\affiliation{Physics Institute, University of S\~{a}o Paulo, 05508-090, S\~{a}o Paulo, Brazil}
\author{B.~B.~Manzato}
\affiliation{Physics Institute, University of S\~{a}o Paulo, 05508-090, S\~{a}o Paulo, Brazil}
\author{R.~H.~Maruyama}
\affiliation{Department of Physics and Wright Laboratory, Yale University, New Haven, CT 06520, USA}
\author{R.~J.~Neal}
\affiliation{Department of Physics and Astronomy, University of Sheffield, Sheffield S3 7RH, United Kingdom}
\author{S.~L.~Olsen}
\affiliation{Center for Underground Physics, Institute for Basic Science (IBS), Daejeon 34126, Republic of Korea}
\author{H.~K.~Park}
\affiliation{Department of Accelerator Science, Korea University, Sejong 30019, Republic of Korea}
\author{H.~S.~Park}
\affiliation{Korea Research Institute of Standards and Science, Daejeon 34113, Republic of Korea}
\author{J.~C.~Park}
\affiliation{Department of Physics and IQS, Chungnam National University, Daejeon 34134, Republic of Korea}
\author{J.~S.~Park}
\affiliation{Department of Physics, Kyungpook National University, Daegu 41566, Republic of Korea}
\author{K.~S.~Park}
\affiliation{Center for Underground Physics, Institute for Basic Science (IBS), Daejeon 34126, Republic of Korea}
\author{K.~Park}
\affiliation{Center for Underground Physics, Institute for Basic Science (IBS), Daejeon 34126, Republic of Korea}
\author{S.~D.~Park}
\affiliation{Department of Physics, Kyungpook National University, Daegu 41566, Republic of Korea}
\author{R.~L.~C.~Pitta}
\affiliation{Physics Institute, University of S\~{a}o Paulo, 05508-090, S\~{a}o Paulo, Brazil}
\author{H.~Prihtiadi}
\affiliation{Department of Physics, Universitas Negeri Malang, Malang 65145, Indonesia}
\author{S.~J.~Ra}
\affiliation{Center for Underground Physics, Institute for Basic Science (IBS), Daejeon 34126, Republic of Korea}
\author{C.~Rott}
\affiliation{Department of Physics and Astronomy, University of Utah, Salt Lake City, UT 84112, USA}
\author{K.~A.~Shin}
\affiliation{Center for Underground Physics, Institute for Basic Science (IBS), Daejeon 34126, Republic of Korea}
\author{D.~F.~F.~S. Cavalcante}
\affiliation{Physics Institute, University of S\~{a}o Paulo, 05508-090, S\~{a}o Paulo, Brazil}
\author{M.~K.~Son}
\affiliation{Department of Physics and IQS, Chungnam National University, Daejeon 34134, Republic of Korea}
\author{N.~J.~C.~Spooner}
\affiliation{Department of Physics and Astronomy, University of Sheffield, Sheffield S3 7RH, United Kingdom}
\author{L.~T.~Truc}
\affiliation{Department of Physics, Kyungpook National University, Daegu 41566, Republic of Korea}
\author{L.~Yang}
\affiliation{Department of Physics, University of California San Diego, La Jolla, CA 92093, USA}
\author{G.~H.~Yu}
\affiliation{Center for Underground Physics, Institute for Basic Science (IBS), Daejeon 34126, Republic of Korea}
\collaboration{COSINE-100 Collaboration}

\author{J.~Amar{\'e}}
\affiliation{Centro de Astropart\'{\i}culas y F\'{\i}sica de Altas Energ\'{\i}as (CAPA), Universidad de Zaragoza, Pedro Cerbuna 12, 50009 Zaragoza, Spain} 
\affiliation{Laboratorio Subterr\'aneo de Canfranc, Paseo de los Ayerbe s.n., 22880 Canfranc Estaci\'on, Huesca, Spain}
 \author{J.~Apilluelo}
 \affiliation{Centro de Astropart\'{\i}culas y F\'{\i}sica de Altas Energ\'{\i}as (CAPA), Universidad de Zaragoza, Pedro Cerbuna 12, 50009 Zaragoza, Spain} 
\affiliation{Laboratorio Subterr\'aneo de Canfranc, Paseo de los Ayerbe s.n., 22880 Canfranc Estaci\'on, Huesca, Spain}
 \author{S.~Cebri{\'an}}
 \affiliation{Centro de Astropart\'{\i}culas y F\'{\i}sica de Altas Energ\'{\i}as (CAPA), Universidad de Zaragoza, Pedro Cerbuna 12, 50009 Zaragoza, Spain} 
\affiliation{Laboratorio Subterr\'aneo de Canfranc, Paseo de los Ayerbe s.n., 22880 Canfranc Estaci\'on, Huesca, Spain}
\author{D.~Cintas}
\affiliation{Centro de Astropart\'{\i}culas y F\'{\i}sica de Altas Energ\'{\i}as (CAPA), Universidad de Zaragoza, Pedro Cerbuna 12, 50009 Zaragoza, Spain} 
\affiliation{Laboratorio Subterr\'aneo de Canfranc, Paseo de los Ayerbe s.n., 22880 Canfranc Estaci\'on, Huesca, Spain}
 \author{I.~Coarasa}
 \affiliation{Centro de Astropart\'{\i}culas y F\'{\i}sica de Altas Energ\'{\i}as (CAPA), Universidad de Zaragoza, Pedro Cerbuna 12, 50009 Zaragoza, Spain} 
\affiliation{Laboratorio Subterr\'aneo de Canfranc, Paseo de los Ayerbe s.n., 22880 Canfranc Estaci\'on, Huesca, Spain}
\author{E.~Garc\'{\i}a}
\affiliation{Centro de Astropart\'{\i}culas y F\'{\i}sica de Altas Energ\'{\i}as (CAPA), Universidad de Zaragoza, Pedro Cerbuna 12, 50009 Zaragoza, Spain} 
\affiliation{Laboratorio Subterr\'aneo de Canfranc, Paseo de los Ayerbe s.n., 22880 Canfranc Estaci\'on, Huesca, Spain}
\author{M.~Mart\'{\i}nez}
\affiliation{Centro de Astropart\'{\i}culas y F\'{\i}sica de Altas Energ\'{\i}as (CAPA), Universidad de Zaragoza, Pedro Cerbuna 12, 50009 Zaragoza, Spain} 
\affiliation{Laboratorio Subterr\'aneo de Canfranc, Paseo de los Ayerbe s.n., 22880 Canfranc Estaci\'on, Huesca, Spain}
\author{Y.~Ortigoza}
\affiliation{Centro de Astropart\'{\i}culas y F\'{\i}sica de Altas Energ\'{\i}as (CAPA), Universidad de Zaragoza, Pedro Cerbuna 12, 50009 Zaragoza, Spain} 
\affiliation{Laboratorio Subterr\'aneo de Canfranc, Paseo de los Ayerbe s.n., 22880 Canfranc Estaci\'on, Huesca, Spain}
\affiliation{Escuela Universitaria Polit\'ecnica de La Almunia de Do\~{n}a Godina (EUPLA), Universidad de Zaragoza, Calle Mayor 5, La Almunia de Do\~{n}a Godina, 50100 Zaragoza, Spain}
\author{A.~Ortiz~de~Sol{\'o}rzano}
\affiliation{Centro de Astropart\'{\i}culas y F\'{\i}sica de Altas Energ\'{\i}as (CAPA), Universidad de Zaragoza, Pedro Cerbuna 12, 50009 Zaragoza, Spain} 
\affiliation{Laboratorio Subterr\'aneo de Canfranc, Paseo de los Ayerbe s.n., 22880 Canfranc Estaci\'on, Huesca, Spain}
\author{T.~Pardo}
\affiliation{Centro de Astropart\'{\i}culas y F\'{\i}sica de Altas Energ\'{\i}as (CAPA), Universidad de Zaragoza, Pedro Cerbuna 12, 50009 Zaragoza, Spain} 
\affiliation{Laboratorio Subterr\'aneo de Canfranc, Paseo de los Ayerbe s.n., 22880 Canfranc Estaci\'on, Huesca, Spain}
\author{J.~Puimed{\'o}n}
\affiliation{Centro de Astropart\'{\i}culas y F\'{\i}sica de Altas Energ\'{\i}as (CAPA), Universidad de Zaragoza, Pedro Cerbuna 12, 50009 Zaragoza, Spain} 
\affiliation{Laboratorio Subterr\'aneo de Canfranc, Paseo de los Ayerbe s.n., 22880 Canfranc Estaci\'on, Huesca, Spain}
\author{M.~L.~Sarsa}
\affiliation{Centro de Astropart\'{\i}culas y F\'{\i}sica de Altas Energ\'{\i}as (CAPA), Universidad de Zaragoza, Pedro Cerbuna 12, 50009 Zaragoza, Spain} 
\affiliation{Laboratorio Subterr\'aneo de Canfranc, Paseo de los Ayerbe s.n., 22880 Canfranc Estaci\'on, Huesca, Spain}
\author{C.~Seoane}
\affiliation{Centro de Astropart\'{\i}culas y F\'{\i}sica de Altas Energ\'{\i}as (CAPA), Universidad de Zaragoza, Pedro Cerbuna 12, 50009 Zaragoza, Spain} 
\affiliation{Laboratorio Subterr\'aneo de Canfranc, Paseo de los Ayerbe s.n., 22880 Canfranc Estaci\'on, Huesca, Spain}

\collaboration{ANAIS-112 Collaboration}

\date{March 21, 2025}

\begin{abstract}
The annual modulation signal, claimed to be consistent with dark matter as observed by DAMA/LIBRA in a sodium-iodide based detector, has persisted for over two decades. \mbox{COSINE-100} and \mbox{ANAIS-112} were designed to test the claim directly using the same target material. \mbox{COSINE-100}, located at Yangyang Underground Laboratory in South Korea, and \mbox{ANAIS-112}, located at Canfranc Underground Laboratory in Spain, have been taking data since 2016 and 2017, respectively. Each experiment published its respective results independently. In this paper, we present the results of an annual modulation search as a test of the signal observed by DAMA/LIBRA with the first three respective years of data from COSINE-100 and \mbox{ANAIS-112}. Using a Markov Chain Monte Carlo method, we find best fit values for modulation amplitude of $-0.0002 \pm 0.0026$\,cpd/kg/keV in the 1--6\,keV and $0.0021 \pm 0.0028$\,cpd/kg/keV in the 2--6\,keV energy regions. These results are not compatible with DAMA/LIBRA's assertion for their observation of annual modulation at 3.7$\sigma$ and 2.6$\sigma$, respectively. Performing a simple combination of the newly released 6-years datasets from both experiments find values consistent with no modulation at $0.0005 \pm 0.0019$\,cpd/kg/keV in the 1--6\,keV and $0.0027 \pm 0.0021$\,cpd/kg/keV in the 2--6\,keV energy regions with 4.7$\sigma$ and 3.5$\sigma$ respective exclusions of the DAMA/LIBRA signal.

\end{abstract}

\maketitle




The energy content of the Universe is expected from cosmological observations to be 27\% dark matter as unaccounted for by standard model particle physics~\cite{Planck_27percent}. Many theories exist that could explain the nature of a dark matter particle. The Weakly Interacting Massive Particle (WIMP) is particularly compelling as it fulfills strong theory motivations and, given controlled uncertainties from astronomical input, could be realized by direct detection experiments alone~\cite{Theory_waningWIMP_2018}.

One way to search for WIMPs and WIMP-like particles is to look for their scattering off of atomic nuclei~\cite{detector_scattering}. In addition, we can look for an annual modulation of this signature induced by the variation in their scattering rate due to the relative motion of the Sun and the Earth~\cite{annual_mod_mom}.

The DAMA/NaI collaboration first claimed observing an annual modulation in their data in 1997~\cite{DAMA_NaI}. Since then, they have mounted a new experiment, DAMA/LIBRA, which now reports a modulation with a significance of 12.9$\sigma$~\cite{DAMA2020}. Their results contradict other direct detection dark matter experiments in the most conventional scenarios for the dark matter particle and halo models~\cite{Schumann_2019}, as well as the energy dependence of the signal expected by the WIMP spin-independent isospin conserving case in the standard halo model~\cite{selfimply}. Attempts to explain these contradictions with alternative dark matter models or detector effects have not been successful~\cite{DAMA2020, noRole_mu_nu}. 

COSINE-100~\cite{COSINEinitperform} and ANAIS-112~\cite{ANAIS1yr} are two experiments explicitly designed to test DAMA by using the same target material, NaI(Tl), which removes all dependencies in the particle and halo model that affect the comparison between experiments using different target nuclei. COSINE-100 and ANAIS-112 crystals were grown by the same producer, Alpha Spectra Inc., using similar raw powder materials. Detector design is also very similar in both experiments. The first crystals were produced in 2012 for the ANAIS experiment. 

In addition to using the same target material, good control of systematics in the calibration of the detectors within the region of interest is mandatory. The most relevant caveat associated with the response of the NaI(Tl) detectors comes from the uncertainties in the knowledge of the quenching factors~(QFs) for the scintillation of nuclear recoils in NaI(Tl), the ratio of light yields from a nuclear recoil with respect to an electron recoil of the same energy.
Thought to be an intrinsic property of the material, the sodium and iodine QFs generally have been found to be consistent across many independent measurements for crystals grown from various powders and techniques, but with a relatively high dispersion. In addition, the most recent measurements show a steady decrease at low recoil energies within the region of interest for the testing of the DAMA/LIBRA result~\cite{Joo2019, Bignell2021, Cintas_QF, COSINE_QF}. Recently, these QFs for sodium recoils have been found to vary with the calibration methods applied~\cite{Cintas_QF}, which could explain some of the dispersion in the values obtained by the different measurements. However, the QFs reported by DAMA/LIBRA are assumed by the collaboration to be energy independent, and the values are significantly higher~\cite{DAMA_QF}. 

At present, it cannot be discarded that differences in the QFs could be found for different crystals, depending on the impurities content, crystalline properties, Tallium doping, etc. This being the case, the comparison between COSINE-100, ANAIS-112 and DAMA/LIBRA would require a good knowledge of the QFs for sodium and iodine recoils in all the experiments. COSINE-100 and ANAIS-112 have carried out dedicated measurements of the QFs for their crystals~\cite{COSINE_QF,Cintas_QF}, and address the analysis of this discrepancy in their recent 6-year dataset publications~\cite{ANAIS6yr,COSINE_full}.

%
The COSINE-100 experiment collected physics data between October 16, 2016 and March 14, 2023 at the Yangyang underground laboratory (Y2L) in Korea. COSINE-100 consisted of five low background NaI(Tl) detectors for a total active mass of \SI{61.3}{kg}. The detectors were shielded by \SI{700}{m} rock overburden (1800 m.w.e), \SI{2200}{liters} of liquid scintillator, copper, lead, and plastic scintillator muon veto panels. The liquid scintillator is an additional veto at low energies, allowing COSINE-100 to conduct dark matter searches between 2--3\,cpd/kg/keV background rates in the energy region of interest~\cite{liq_scin}. The COSINE-100 detector is described in~\cite{COSINEG4}. \mbox{COSINE-100's} signal selection efficiency is 85\% at \SI{1.0}{keV} and approaches unity at \SI{2}{keV}~\cite{3yrresultCOSINE}.


ANAIS-112 began taking data in August 3, 2017 at the Laboratorio Subterr\'aneo de Canfranc (LSC) in Spain, under a rock overburden of 2450 m.w.e. \mbox{ANAIS-112} will continue in operation until the end of 2025.  
\mbox{ANAIS-112} consists of nine NaI(Tl) crystals, \SI{12.5}{kg} each, for an active mass of \SI{112.5}{kg}. 
 The shielding is composed of archaeological lead, low-activity lead, an anti-radon box, active muon vetoes, polyethylene bricks, and water tanks~\cite{ANAIS1yr, Coarasa_2023}. The background rate in the region of interest amounts to 2.0--4.7\,cpd/kg/keV, depending on the crystal, and  efficiencies around 95\% in the 1--6\,keV energy region have been checked with $^{109}$Cd and $^{252}$Cf periodical calibrations~\cite{ANAIS6yr}.


With ANAIS-112's higher active mass and \mbox{COSINE-100's} lower background rate because of the liquid scintillator veto, the two experiments have similar sensitivities.
ANAIS-112 and COSINE-100 have published results from the first three years of data, corresponding to live times of 2.78~years and 2.58~years, respectively~\cite{ANAIS3yr_update, 3yrresultCOSINE}. 
Both experiments find no modulation to support the DAMA/LIBRA dark matter claim, and their results are displayed in Table~\ref{table:results}. 
The datasets for both experiments' dark matter annual modulation search are made available open access through Origin Excellence Cluster's Dark Matter Data Center~\cite{DMDC}.
Recently, COSINE-100 and ANAIS-112 have released  modulation searches with datasets corresponding to 5.85~years~\cite{COSINE_full} and 5.56~years live time~\cite{ANAIS6yr}, respectively. Both results are compatible with the no modulation case. 

Here, we present a combined analysis of the first three years of data from \mbox{ANAIS-112} and \mbox{COSINE-100} using all the information available on the time distribution of the rates, efficiencies, and background modeling for both experiments and compare the Bayesian and frequentist approaches. Additionally, we compare these results with those from a simple (weighted) combination of the best-fit values, finding them to be compatible. We also present the simple combination of the six-year results as an indication of what can be expected when the data are released and analyzed by the methods described here. 





\textit{Computation of Residuals--}--Due to the numerous factors each experiment must individually account for, the best way to directly combine datasets is to compute the residuals of each crystal detector for each experiment and then, perform a combined fit to those data. 

Each experiment has provided its measured event rate with the respective crystal detector live times and efficiencies. For the purposes of the modulation search, we are interested in the single-hit selected events as random coincidences of dark matter and background signals are negligible.
We have seen from Ref.~\cite{Hafizh_paper} that an incomplete description of time-dependent backgrounds can artificially induce modulation signals. As such, each experiment has conducted a thorough investigation of time-dependent backgrounds~\cite{COSINEbkgd,ANAIS_bkgd2021}. 
Included in these models are crucial information about the radioactive background content of the experiments where, in the ROI, the dominant components are $^{210}$Pb on the crystal surface and, in the crystal bulk, $^3$H and $^{40}$K.

\textit{ANAIS-112 Background Model--}--ANAIS-112 has explored different techniques for the background modeling in the annual modulation analysis for the 3-year dataset, including a single exponential decay and Monte-Carlo (MC) probability density function (PDF) models, both added to a constant component~\cite{ANAIS3yr, Ivan}. The MC PDF background model used by ANAIS is built by the addition of all the background contributions identified in the ANAIS-112 background model developed in Ref.~\cite{ANAIS_1st_bkgd} and updated with the three-year data and improved analysis methods developed in Refs.~\cite{ANAIS_bkgd2021,ANAIS3yr}. This MC PDF takes into account for each energy region the evolution in time of each contribution, according to the lifetimes of each parent isotope. Further updating is in progress using the available six-year data.

Though the MC PDF model offers one fewer nuisance parameter, the alternative single exponential decay was found to be a good fit with \mbox{$\chi^2\approx$ 129/107} and 115/107 in the 1--6 and 2--6\,keV regions, respectively, for the modulation hypothesis~\cite{ANAIS3yr}. The sensitivity achieved by the single exponential model matches well that of the MC~PDF's. The single exponential model used in ANAIS is:
\begin{equation}
    \phi_{bkgd}(t_i) = (1-f)+fe^{-t_i/\tau} 
    \label{eqn:single_exp}
\end{equation}
where the two free parameters are the overall scaling factor, $f$, and time-dependent backgrounds effective decay time constant, $\tau$. The exponential function is used to model the \mbox{ANAIS-112} backgrounds in this combined analysis, as it is straightforward to implement and yields statistically equivalent results. The analysis of the ANAIS-112 3-year data following this model yields a non-statistically significant difference with the modulation search results in Ref.~\cite{ANAIS3yr_update} and observes no change in sensitivity. 

\textit{COSINE-100 Background Model--}--COSINE-100 time-dependent background model uses a sum of eight exponential decays, 
resulting from $^3$H, $^{22}$Na, $^{109}$Cd, $^{210}$Pb in the bulk of the NaI crystal, $^{210}$Pb on near-by surfaces, $^{113}$Sn, $^{121m}$Te, and $^{127m}$Te, plus a constant ``flat'' term which accounts for the long-lived isotopes:
%
\begin{equation}
R_i\left(t \mid \alpha_i, \beta_i\right)=\alpha_i+\sum_{k=1}^{N_{b k g d}} \beta_{0, k}^i e^{-\lambda_k t},
\label{eqn:cos_bkgd}
\end{equation}
where $\alpha_i$ is the flat component of the $i^{th}$ crystal, $\beta_{0,k}$ is the initial rate 
at $t=0$ of the $k^{th}$ short-lived radioisotope, and $\lambda_k$ is the decay constant of that component. The thorough modeling of these components is described in~\cite{COSINEbkgd2024, COSINEbkgd, COSINEG4}.
It is worth noting that the surface-$^{210}$Pb component has an ``effective'' half-life of a computed $33.8 \pm 8$~years due to it being replenished from the $^{222}$Rn decay chain~\cite{COSINEbkgd2024}. 

Additionally, this study uses the updated background model from COSINE-100 in Ref.~\cite{COSINEbkgd2024}. We re-analyze the 3-year data and find a slight, non-statistically significant, difference with the previous COSINE-100 \mbox{3-year} modulation search results in Ref.~\cite{3yrresultCOSINE}, and observe no change in sensitivity (see Table~\ref{table:results}).

\begin{figure}[!hbt]
    \begin{center}
    \includegraphics[width=1\columnwidth]{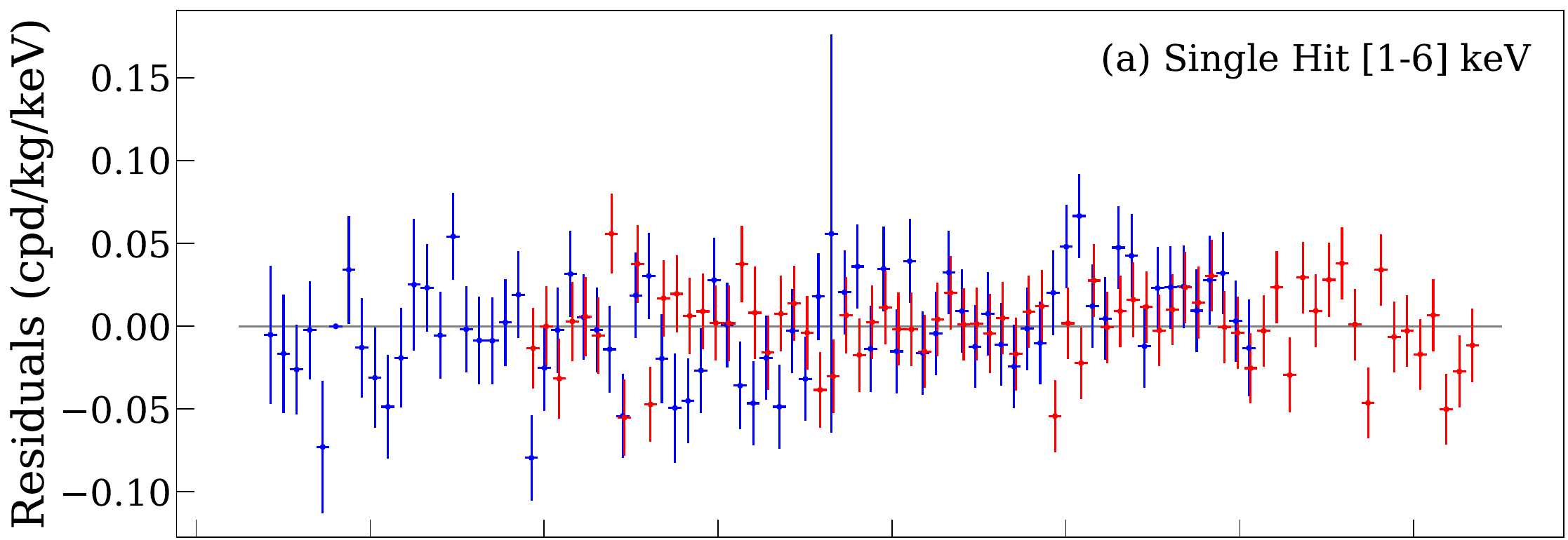}	    
    \includegraphics[width=1\columnwidth]{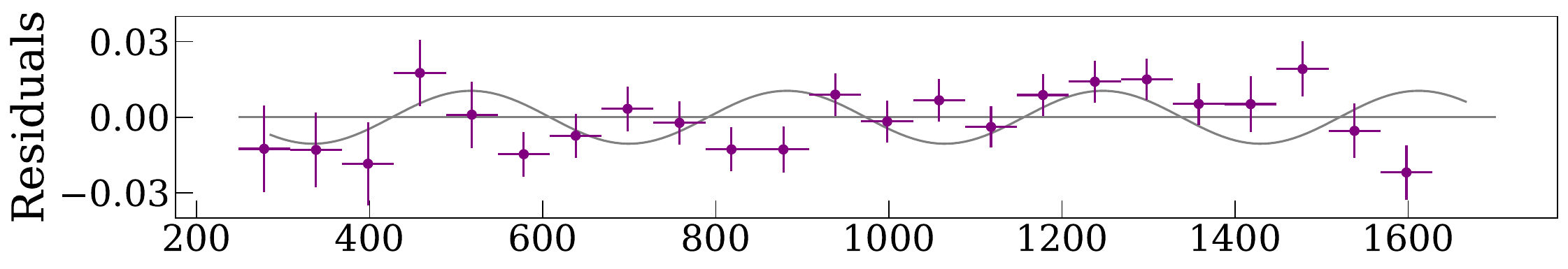}
    \includegraphics[width=1\columnwidth]{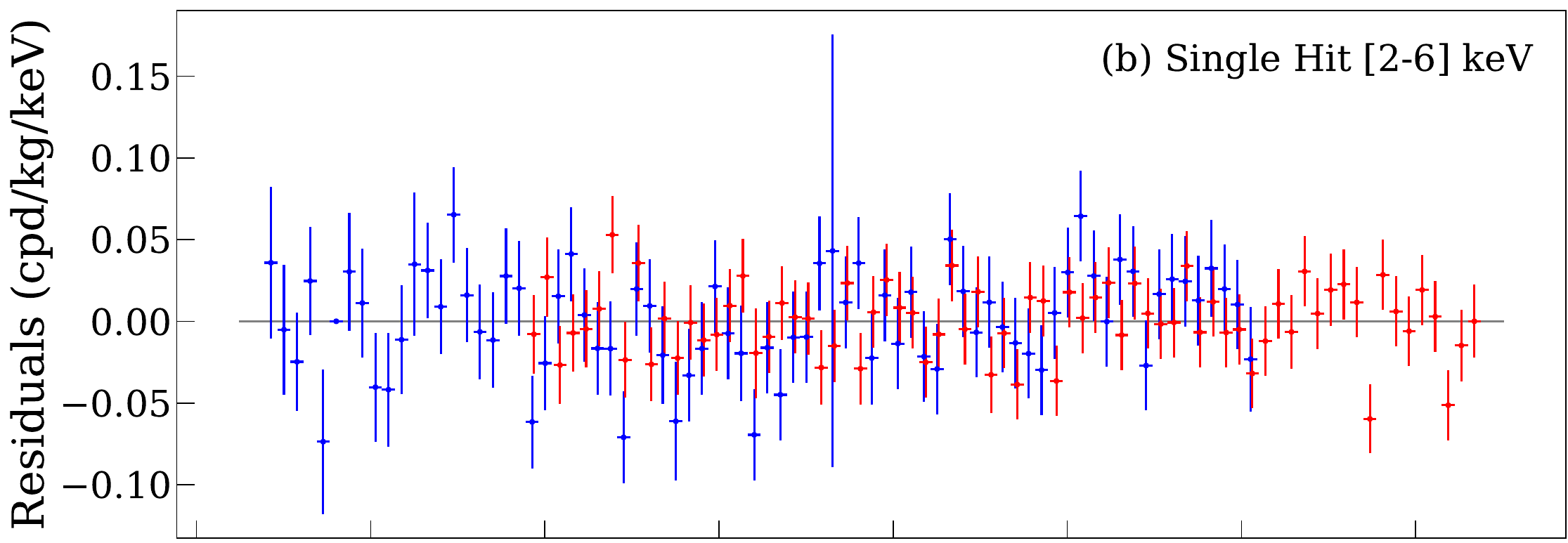}
    \includegraphics[width=1\columnwidth]{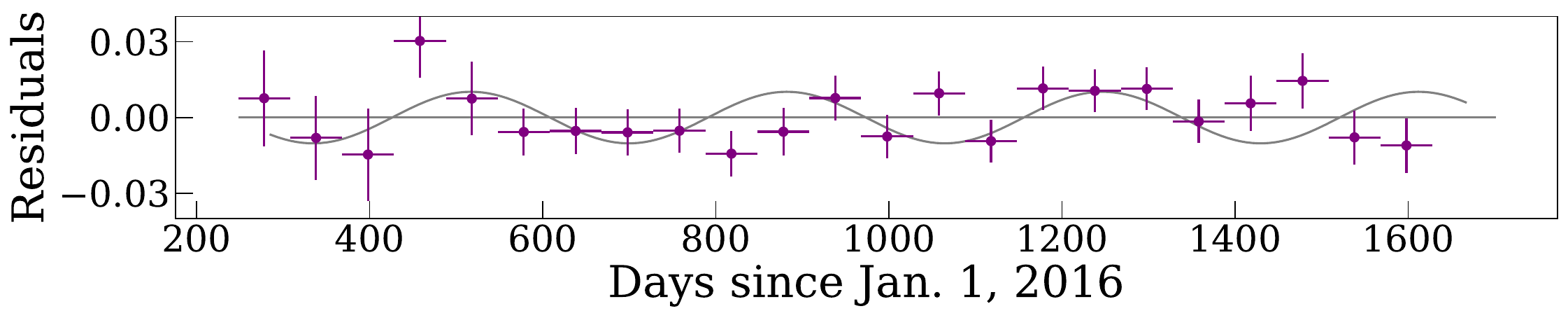}
	    \caption{The background subtracted residuals for \mbox{COSINE-100}~(blue) and \mbox{ANAIS-112}~(red),for a total exposure of 485\,kg$\cdot$yr in energy ranges of (a) 1--6 and (b) 2--6\,keV. The subplots show their combined data in time bins of 2 months (purple) where a grey curve is drawn to visually compare the DAMA modulation signal. }
	    \label{fig:combined_residuals}
	\end{center}
\end{figure}

\textit{Annual Modulation Search--}--Residuals retrieved from subtracting modeled backgrounds are plotted together in Fig.~\ref{fig:combined_residuals}. We used 15-day bins for both experiments, January~1,~2016 as the common start date for $t=0$, reaching a total exposure of 485\,kg$\cdot$yr. 
Residuals from COSINE-100 and ANAIS-112 are simultaneously fit for the annual modulation signal. In the standard halo dark matter model, the rate of scattering events in the background subtracted residuals is expected to be: 
\begin{equation}
    R(t) = S_m \cos(\omega(t-t_0)),
    \label{eqn:modulation}
\end{equation}
where $\omega = \frac{2\pi}{T}$ for the period of $T=$ 1~year, $t_0$ is the phase (fixed to June~2), and $S_m$ is a modulation amplitude. 
DAMA/LIBRA reports a modulation amplitude of \mbox{$S_m=0.0105\pm0.0011$ ($0.0102\pm0.0008$)\,cpd/kg/keV} in the 1--6~(2--6\,keV) energy region~\cite{DAMA2020}. It is also possible to simultaneously fit the background models of the respective experiments along the modulation function, bypassing the need to compute residuals, which retrieves equivalent results.



\textit{Bayesian vs.~Frequentist--}--We carried out both a $\chi^2$ minimization of the Least-Squares (i.e.~frequentist) method and Markov Chain Monte Carlo (MCMC, i.e.~Bayesian) analysis to fit the data shown in Fig.~\ref{fig:combined_residuals} with Equation~\ref{eqn:modulation}. 
While each analysis method has its benefits and drawbacks, we illustrate here that both methods give consistent results for the annual modulation search using the COSINE-100 and ANAIS-112 datasets. The COSINE collaboration uses a Bayesian approach to modulation search, while the ANAIS collaboration has performed a frequentist method using $\chi^2$ minimization~\cite{3yrresultCOSINE, ANAIS3yr_update}.

The annual modulation results of \mbox{COSINE-100}, \mbox{ANAIS-112}, their combined search, and the comparison of statistical methods are summarized in Table~\ref{table:results}. The previously published values of COSINE-100~\cite{3yrresultCOSINE}, ANAIS-112~\cite{ANAIS3yr_update}, and DAMA/LIBRA~\cite{DAMA2020} are displayed in Table~\ref{table:compare} for comparison. Despite the differences in background models with the previously published searches, the sensitivities found for COSINE-100 and ANAIS-112 by this study are unchanged. 

\begin{figure}[]
    \begin{center}
	    \includegraphics[width=0.95\columnwidth]{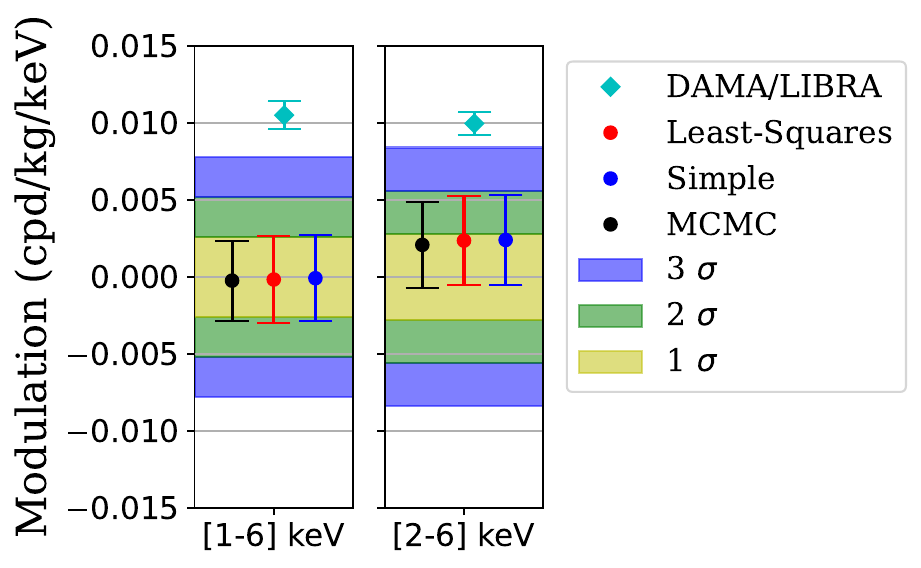}
	    \caption{ Best-fit amplitudes for the \mbox{1--6} and 2--6\,keV regions where the colored bands show the combined sensitivity. There is remarkable compatibility between results of the MCMC, Least-Squares fit, as well as the simple combination of the COSINE-100 and ANAIS-112 independent results.  }
	    \label{fig:results}
	\end{center}
\end{figure}

Fig.~\ref{fig:results} displays the results for the analysis performed via the Least-Squares and MCMC, as well as the simple combination of the two independent experiments' results. 
The simultaneous fit of the 3-year datasets from \mbox{COSINE-100} and \mbox{ANAIS-112} to Equation~\ref{eqn:modulation} finds no modulation, with MCMC best-fit amplitudes at \mbox{$-0.0002 \pm 0.0026$\,cpd/kg/keV} and \mbox{$0.0021 \pm 0.0028$\,cpd/kg/keV}, for \mbox{1--6} and \mbox{2--6}\,keV, respectively. These values are compatible with the amplitudes found by the Least-Squares fit as well as the simple combination of the COSINE-100 and ANAIS-112 independent results.

\begin{table}[t]
\setlength{\tabcolsep}{5pt} 
\renewcommand{\arraystretch}{1.2} 
\caption{
Comparison of annual modulation search results of \mbox{COSINE-100}, \mbox{ANAIS-112}, and their combined search with the 3-year datasets found by this study. These values reflect the decision of background model and include our re-analysis for the COSINE-100 3-year dataset using the updated background model~\cite{COSINEbkgd2024}. The Least-Squares (LS) frequentist method is shown beside the MCMC Bayesian method. All results show the modulation amplitude in\,cpd/kg/keV for the frequency and phase-fixed scenario.}\label{table:results}
\centering
\sisetup{separate-uncertainty = true}
\begin{tabular}{lcc}
    \hline \hline 
    Configuration & {1--6\,keV} & {2--6\,keV} \\ \hline
COSINE-100 (this)   & $\phantom{-}0.0052\pm0.0042$   & $\phantom{-}0.0067\pm0.0047$   \\
ANAIS-112 (this)    & $-0.0040\pm0.0037$  & $-0.0002\pm0.0037$  \\
Combined MCMC   & $-0.0002\pm0.0026$  & $\phantom{-}0.0021\pm0.0028$   \\
Combined LS     & $-0.0002\pm0.0028$  & $\phantom{-}0.0023\pm0.0029$   \\
Combined Simple & $-0.0001\pm0.0028$  & $\phantom{-}0.0024\pm0.0029$   \\ \hline \hline
\end{tabular}
\end{table}

\begin{table}[t]
\setlength{\tabcolsep}{6pt} 
\renewcommand{\arraystretch}{1.2} 
\caption{
Comparison of annual modulation search results from \mbox{COSINE-100}~\cite{3yrresultCOSINE}, \mbox{ANAIS-112}~\cite{ANAIS3yr_update}, and DAMA/LIBRA~\cite{DAMA2020}. All results show the modulation amplitude in\,cpd/kg/keV for the frequency and phase-fixed scenario.}\label{table:compare}
\centering
\sisetup{separate-uncertainty = true}
\begin{tabular}{lcc}
    \hline \hline 
    Configuration & {1--6\,keV} & {2--6\,keV} \\ \hline
COSINE-100~\cite{3yrresultCOSINE}   & $\phantom{-}0.0067\pm0.0042$   & $0.0050\pm0.0047$   \\
ANAIS-112~\cite{ANAIS3yr_update}    & $-0.0013\pm0.0037$  & $0.0031\pm0.0037$  \\
DAMA/LIBRA~\cite{DAMA2020}   & $\phantom{-}0.0105\pm0.0011$   & $0.0102\pm0.0008$ \\ \hline \hline
\end{tabular}
\end{table}
\textit{Method Verification through Simulations--}--A comprehensive investigation for the biases of the analysis is conducted by employing a pseudo study to quantify the impact of systematics. COSINE-100 and ANAIS-112 have each performed their own pseudo studies and found no relevant biases~\cite{3yrresultCOSINE,ANAIS3yr}. Since the method for fitting the modulation to residuals is specific to this study, we present the results for the bias investigation by pseudo study for this method. 

We simulated the data for each crystal in both experiments by taking the COSINE and ANAIS background models described in Equations~\ref{eqn:single_exp} and \ref{eqn:cos_bkgd}, inject various amplitudes for the dark matter signal, and vary them with a Poisson distribution to mimic a counting experiment. 

Data generated from the background model produces a simulated experiment for the null hypothesis case without the dark matter signal. 
We build ensembles by  injecting  the dark matter signal with varying $S_m$ amplitudes of $\pm 0.0025$, $\pm 0.005$, $\pm 0.0075$, and the DAMA/LIBRA signal, $\pm 0.0105$\,cpd/kg/keV, with fixed phase~(June 2) and period~(1~year). Including the no modulation case, this gives a total of 9 ensembles. 
 Approximately 10,000 pseudo-experiments were generated per ensemble to quantify potential biases, defined as the difference between the injected signal and the fit result. For all ensembles, mean biases were found to be negligible. 

\begin{figure}[t]
    \begin{center}
	    \includegraphics[width=0.45\columnwidth]{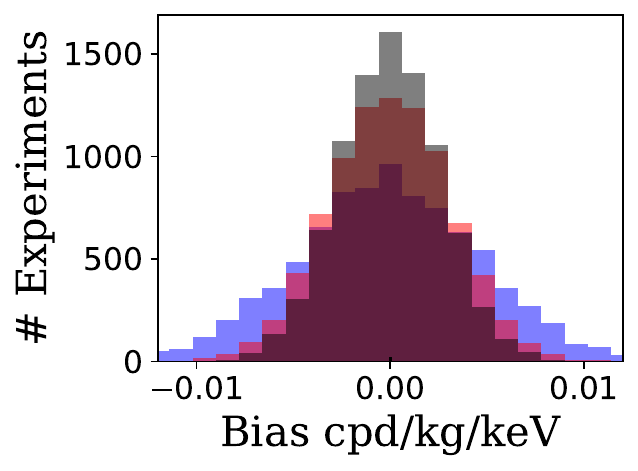}	    \includegraphics[width=0.45\columnwidth]{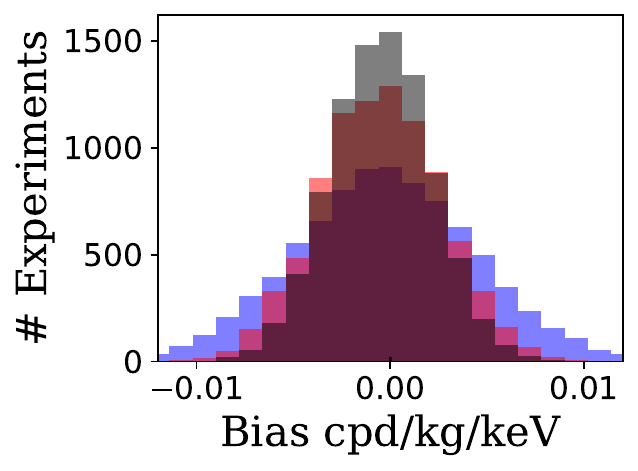}
        \includegraphics[width=0.45\columnwidth]{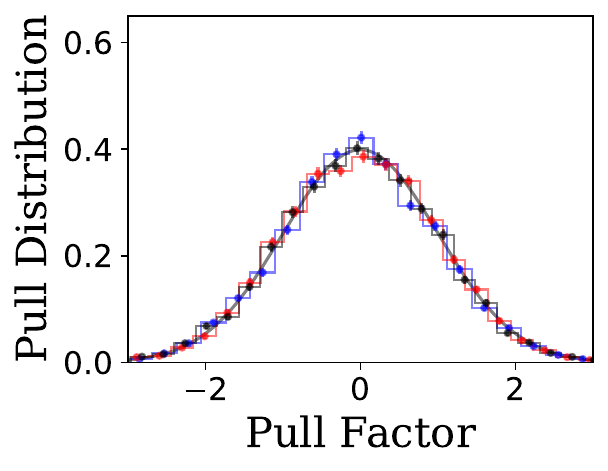}
        \includegraphics[width=0.45\columnwidth]{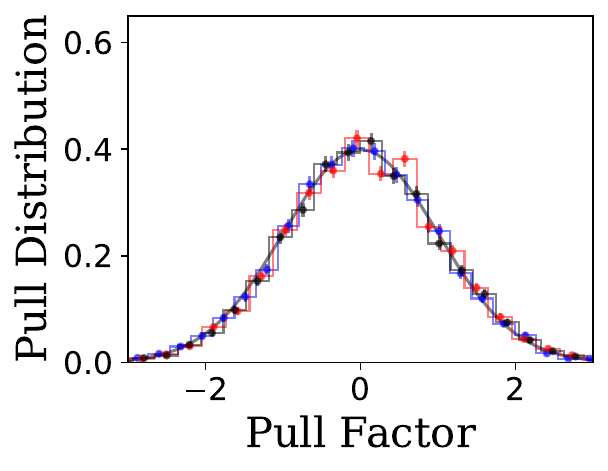}
	    \caption{
        (Top) Distribution of the bias between best-fit and injected modulation amplitudes in the 1--6\,keV region for null hypothesis (top left) and DAMA modulation (top right) cases.  (Bottom) Pull distributions for the 1--6\,keV (bottom left) and 2--6\,keV region (bottom right) where all are compatible with $\mu=0$ and $\sigma=1$ (grey curve). The blue, red, and black shaded regions correspond to \mbox{COSINE-100}, \mbox{ANAIS-112}, and their combined analysis, respectively. See Table~\ref{table:bias} for reported numerical values for the mean biases. 
       }
	    \label{fig:mean_bias}
	\end{center}
\end{figure}

Fig.~\ref{fig:mean_bias} shows the bias distribution for the 10,000 experiments of the null hypothesis and the DAMA/LIBRA modulation cases in the top panel. The bottom panel shows the pull distribution which is computed by dividing the bias by the fit uncertainty and characterizes the error bias. These are compatible with the standard normal distribution and reflect that the sensitivity of the experiment(s) are therefore compatible with the spread of the pseudo modulation results. Table~\ref{table:bias} lists this value in the last column and shows the corresponding results for the mean bias for the null hypothesis and DAMA/LIBRA modulation cases. 

\begin{table}[t]
\setlength{\tabcolsep}{5pt} 
\renewcommand{\arraystretch}{1.2} 
\caption{Bias in $\times 10^{-4}$\,cpd/kg/keV (defined as the difference of the injected signal and the fit result), found by the fitting procedures derived from simulation for the no modulation case and the DAMA/LIBRA modulation in the 1--6\,keV and 2--6\,keV energy regions.  The bias listed for \mbox{ANAIS-112} corresponds to the background model from Equation~\ref{eqn:single_exp} as reported in Ref.~\cite{ANAIS3yr} and the COSINE-100 bias uses the updated background model from Ref.~\cite{COSINEbkgd2024}. The last column shows the standard deviation in $\times 10^{-4}$\,cpd/kg/keV of the pseudo study modulations.}\label{table:bias}
\begin{tabular}{cccc}
\hline \hline
1--6\,keV      & \begin{tabular}[c]{@{}c@{}}Bias [$S_m=0$]    \end{tabular} & \begin{tabular}[c]{@{}c@{}}Bias [DAMA $S_m$]  \end{tabular} & $\sigma_{Sm}$ \\ \hline
COSINE-100      & $-0.2 \pm 0.4$                                                  & $-3.9 \pm 0.3$     &   $47 \pm 4$    \\ 
ANAIS-112 & $-0.4 \pm 0.2$                                                & $-4.7 \pm 0.2$ & $35 \pm 3$  \\ 
Combined  & $-0.3 \pm 0.2$                                                  & $-3.8 \pm 0.2$ & $28 \pm 2$ \\  \hline \hline
2--6\,keV        &  & \\ \hline
COSINE-100      & $-0.3 \pm 0.4$                                                  & $-4.3 \pm 0.4$ & $52 \pm 5$\\ 
ANAIS-112 & $-0.3 \pm 0.2$                                                & $-4.7 \pm 0.3$  & $35 \pm 3$ \\ 
Combined  & $-0.2 \pm 0.2$                                                  & $-4.4 \pm 0.2$  & $29 \pm 3$ \\ \hline \hline

\end{tabular}
\end{table}

\textit{Towards 1~\lowercase{ton}$\cdot$\lowercase{yr} Combined Exposure--}--\mbox{COSINE-100} has recently released its search for annual modulation with the full dataset, utilizing an exposure of 358\,kg$\cdot$yr~\cite{COSINE_full}. Likewise, ANAIS-112 has recently released their 6-year dataset with an exposure of 626\,kg$\cdot$yr~\cite{ANAIS6yr}.
The procedures of this paper have shown that the data from the two experiments are compatible and that their results can be combined; the direct combination of results provides a similar value to the thorough analysis which can utilize either MCMC or $\chi^2$ minimization. 

\begin{figure}[t]
    \begin{center}
\	    \includegraphics[width=0.92\columnwidth]{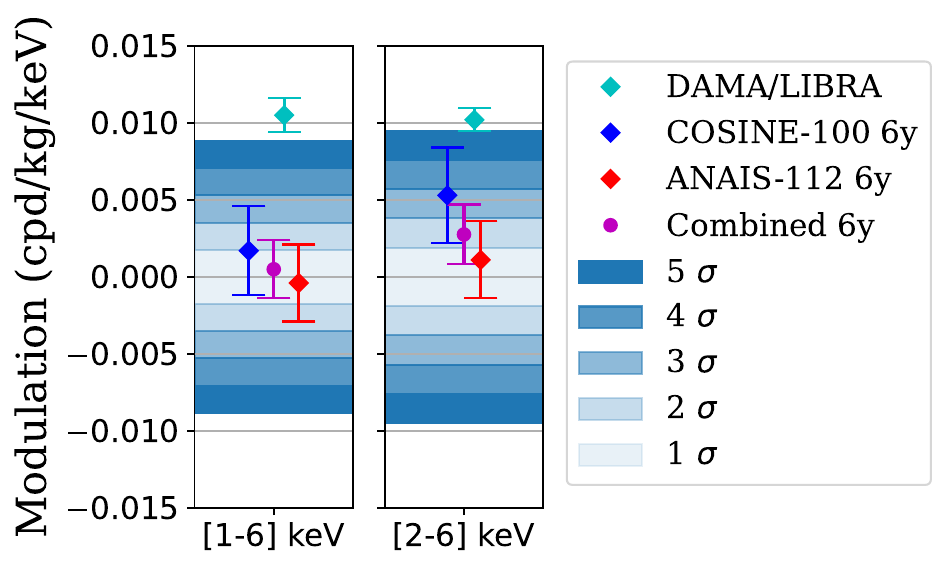}
	    \caption{ Simple combination results of the COSINE-100 full dataset~\cite{COSINE_full} and ANAIS-112 6-year~\cite{ANAIS6yr} annual modulation searches. The colored bands show the sensitivity region for 6-year data from both experiments combined in 1$\sigma$ (lightest blue) to 5$\sigma$ (darkest blue).}
	    \label{fig:6yr}
	\end{center}
\end{figure}

The simple combination for the combined exposure of 984\,kg$\cdot$yr results in modulation values of $0.0005 \pm 0.0019$\,cpd/kg/keV in the 1--6\,keV and \mbox{$0.0027 \pm 0.0019$\,cpd/kg/keV} in the 2--6\,keV energy regions. These values are incompatible with the DAMA/LIBRA signal to significances of 4.7$\sigma$ and 3.5$\sigma$, respectively, as illustrated by Fig.~\ref{fig:6yr}. 
 This result strongly challenges the interpretation of the DAMA/LIBRA modulation in terms of galactic dark matter. A thorough combined analysis will be performed when the full ANAIS-112 dataset is available. 

\textit{Conclusions--}--The methods described in this paper demonstrate the compatibility of the COSINE-100 and ANAIS-112 experiments. The direct combination of the residual rates obtained by the implementation of the respective background models show careful consideration for possible systematics. 
Using the MCMC method, we find best fit values on the combined 3-year datasets from COSINE-100 and ANAIS-112 with modulation amplitude of \mbox{$-0.0002 \pm 0.0026$\,cpd/kg/keV} in the \mbox{1--6\,keV} and \mbox{$0.0021 \pm 0.0028$\,cpd/kg/keV} in the \mbox{2--6\,keV} energy regions. These results are incompatible with DAMA/LIBRA's assertion for their observation of annual modulation at 3.7$\sigma$ and 2.6$\sigma$, respectively.

Furthermore, a simple combination of the newly released 6-years datasets from COSINE-100 and ANAIS-112 find values consistent with no modulation at \mbox{$0.0005 \pm 0.0019$\,cpd/kg/keV} in the 1--6\,keV and \mbox{$0.0027 \pm 0.0021$\,cpd/kg/keV} in the 2--6\,keV energy regions with 4.7$\sigma$ and 3.5$\sigma$ respective exclusions of the DAMA/LIBRA signal.


\begin{acknowledgements}
We thank the Korea Hydro and Nuclear Power (KHNP) Company for providing underground laboratory space at Yangyang and the IBS Research Solution Center (RSC) for providing high performance computing resources. 
This work is supported by:  the Institute for Basic Science (IBS) under project code IBS-R016-A1,  NRF-2021R1A2C3010989, NRF-2021R1A2C1013761 and RS-2024-00356960, Republic of Korea;
NSF Grants No. PHY-1913742, United States; 
STFC Grant ST/N000277/1 and ST/K001337/1, United Kingdom;
Grant No. 2021/06743-1, 2022/12002-7 and 2022/13293-5 FAPESP, CAPES Finance Code 001, CNPq 304658/2023-5, Brazil;
UM Internal Grant non-APBN 2025, Indonesia,
financial support by MCIN/AEI/10.13039/501100011033 under grant PID2022-138357NB-C21 and financial support by MCIN/AEI/10.13039/501100011033 under Grants No. PID2022-138357NB-C21 and No. PID2019–104374 GB-I00, and the Gobierno de Aragón, European Social Fund and funds from European Union NextGenerationEU/PRTR (Planes complementarios, Programa de Astrofísica y Física de Altas Energías).
\end{acknowledgements}

\bibliography{reference_master}

\end{document}